\documentclass[final,5p,times,twocolumn]{elsarticle}
 \biboptions{comma,sort&compress}
\usepackage{graphicx}
\usepackage{here}
\usepackage[dvips]{epsfig}
\def\beq{\begin{equation}}
\def\eeq{\end{equation}}
\def\bea{\begin{eqnarray}}
\def\eea{\end{eqnarray}}
\def\bq{\begin{quote}}
\def\eq{\end{quote}}

\def\nnb{\nonumber}
\def\ga{\left(}
\def\dr{\right)}

\def\lrar{\Longrightarrow}
		
\def\nnb{\nonumber}
\def\la{\langle}
\def\ra{\rangle}
\def\nin{\noindent}
\def\ba{\vspace*{-0.2cm}\begin{array}}
\def\ea{\end{array}\vspace*{-0.2cm}}

\def\b{$\bullet~$}
\def\als{\alpha_s}

\def\gg2{ \la\alpha_s G^2 \ra}
\def\gg3{g^3f_{abc}\la G^aG^bG^c \ra}
\def\ggg4{\la\als^2G^4\ra}

\def\beq{\begin{equation}}
\def\enq{\end{equation}}
\def\beqa{\begin{eqnarray}}
\def\enqa{\end{eqnarray}}
\def\nnb{\nonumber}

\def\qq{\lag\bar{q}q\rag}

\def\mix{\lag\bar{q}g\si Gq\rag}

\def\si{\sigma}

\newcommand{\rag}{\rangle}
\newcommand{\lag}{\langle}

\journal{Elsevier}

\begin{document}

\begin{frontmatter}

\title{Revisiting $f_{B}$ and $\overline{m}_b(\overline{m}_b)$ from  HQET spectral sum rules}
 \author[label2]{Stephan Narison}
\address[label2]{Laboratoire Particules et Univers de Montpellier, CNRS-IN2P3, 
Case 070, Place Eug\`ene
Bataillon, 34095 - Montpellier, France.}
   \ead{snarison@yahoo.fr}
\begin{abstract}
\nin
Using recent values of the QCD (non-) perturbative parameters given in Table \ref{tab:param}, we reconsider the extraction of $f_B$ and the on-shell mass $M_b$ from HQET Laplace spectral sum rules known to N2LO PT series and including dimension 7 condensates in the OPE.
We especially study the convergence of the PT series, the effects on ``different spectral sum rules data" of the continuum threshold and subtraction point varied in a larger range than in the existing literature and include in the error an estimate of the N3LO PT series based on a geometric growth of the PT series. We obtain the Renormalization Group Invariant (RGI) {\it universal coupling} : $\hat f_B^\infty=0.416(60)$  GeV$^{3/2}$  in the static limit $M_b \to \infty$ and the physical decay constant including $1/M_b$ corrections: $f_B^{hqet}=199(29)$ MeV.  Using the ratio of sum rules, we obtain,  to order $\alpha_s^2$,  the running mass $\overline{m}_b(\overline{m}_b)=4213(59)$ MeV. The previous results are in good agreement with the ones from QCD spectral sum rules (QSSR) in full QCD to the same order  from the same channel \cite{SNFB12}:  $f_B^{qcd}=206(7)$ MeV and $\overline{m}_b(\overline{m}_b)^{qcd}=4236(69)$ MeV. 
\end{abstract}
\end{frontmatter}
\section{Introduction }
The (pseudo)scalar meson decay constants $f_P$ are of prime interests for understanding the realizations of chiral symmetry in QCD. 
In addition to the well-known values of $f_\pi=130.4(2)$ MeV and $f_K=156.1(9)$ MeV \cite{ROSNER} which control  the light flavour chiral symmetries, it is also desirable to extract the ones of the heavy-light charm and bottom quark systems with high-accuracy. These decay constants are normalized through the matrix element:
\beq
\la 0|J^P_{\bar qQ}(x)|P\ra= f_P M_P^2~,
\label{eq:fp}
\eeq
where: 
\beq
J^P_{\bar qQ}(x)\equiv (m_q+M_Q)\bar q(i\gamma_5)Q~,
\label{eq:current}
\eeq
is the local heavy-light pseudoscalar current;  $q\equiv d,s;~Q\equiv c,b;~P\equiv D_{(s)}, B_{(s)}$and where $f_P$  is related to  the leptonic width:
\beq
\Gamma (P^+\to l^+\nu_l)={G^2_F\over 8\pi}|V_{Qq}|^2f^2_Pm_l^2M_P\ga 1-{m_l^2\over M_P^2}\dr^2~,
\eeq
where $m_l$ is the lepton mass and $|V_{Qq}|$ the CKM mixing angle.  In a recent analysis \cite{SNFB12}, we have revised the extraction of these heavy-light decay constants in full QCD \cite{SNFB12} using QCD spectral sum rules \cite{SVZ,RRY,SNB1,SNB2,SNB3,CK}. Here, we pursue the analysis by revisiting the determination of $f_B$ from HQET spectral sum rules. In so doing, we shall 
explicitly analyze the influence on the results of the subtraction point $\mu$ and of the continuum threshold $t_c$. We shall also use (besides recent determinations of the QCD input parameters) the new precise value of $m_b$ from the $\Upsilon$ sum rules \cite{SNH10}. In addition, we shall re-extract the meson-quark mass difference using HQET sum rules from which we shall deduce the running $b$-quark mass.
\section{HQET preliminaries }
 HQET spectral sum rules have been initially used by Shuryak \cite{SHURYAK} using a non-relativistic \footnote{Some earlier attempt to use a non-relativistic approach for estimating $f_D$ can e.g. be found in \cite{KHOLOPOV}.} version of the NSVZ \cite{NSV2Z} sum rules in the large $M_b$ limit.  Shuryak's sum rule has been applied later on in HQET \cite{HQET} by several authors \cite{HUSSAIN,SNIMET,GROZIN,DOSCH,NEUBERT,ELETSKY,PENIN}. The most important input in the analysis of HQET sum rule is the local heavy-light quark axial-vector current of the full QCD theory which can be expressed as an OPE of the HQET operators $\tilde {\cal O}_n$ in the inverse of the heavy quark mass:
\bea
J^\mu_A (x,M_b)&=& C_b\ga {M_b\over\mu},\alpha_s(\mu)\dr \tilde {J}^\mu_A(x,M_b= \infty)\nnb\\
&&+ \sum_{n=1}C_n\ga {M_b\over\mu},\alpha_s(\mu)\dr {\tilde {\cal O}_n\ga M_b=\infty,\mu\dr\over M_b^n}~,
\eea
where : $\tilde {J}^\mu_A\equiv \bar q\gamma^\mu\gamma^5h_v$ is the quark current in HQET built from a light antiquark field $\bar q$ and a properly normalized heavy quark field $h_v$ \cite{HQET}, $C_{b,n}$ are Wilson coefficients and $M_b$ is the on-shell $b$-quark mass. Using a non-covariant normalization of hadronic states \cite{EICHTEN}, one can  define an universal coupling in the static limit:
\beq
\la 0|\tilde {J}^\mu_A|P(v)\ra ={i\over \sqrt{2}}\tilde{f}_{stat}v^\mu~.
\eeq
The coefficient function $C_b\ga {M_b/ \mu},\alpha_s(\mu)\dr$ is obtained by requiring that 
HQET reproduces the full QCD theory at $\mu=M_b$. It has been obtained to order $\alpha_s$ in \cite{DOSCH} and to order $\alpha_s^2$ in \cite{GROZIN2}. It reads in the $\overline{MS}$-scheme:
\bea
C_b\ga {M_b}\dr&=& 1-{2\over 3}a_s\ga {M_b}\dr+a_s^2\ga {M_b}\dr\Bigg{[} -{1871\over 1729}
-{17\pi^2\over 72}\nnb\\
&&-{\pi^2\over 18}\ln{2}-{11\over 36}\zeta(3)
+n_l\ga {47\over 288}+\pi^2\over 36\dr\Bigg{]}~,
\eea
with : $a_s\equiv \alpha_s/\pi$. The HQET current $\tilde {J}^\mu_A$ acquires anomalous dimension, which reads to \cal{O}$(\alpha_s^2)$ \cite{HQET,JI,BROAD} (in our normalizations) for $n_l$ flavours:
\bea
\gamma&\equiv&\gamma_1~ a_s +\gamma_2~ a_s^2+...,\nnb\\
\gamma_1&=&1~,~~~~~\gamma_2={127\over 72}+{7\pi^2\over 54}-{5\over 36}n_l~. 
\eea
Therefore, the universal coupling scales as:
\beq
\tilde f_{stat}(\mu)={R_b(M_b)\over R_b(\mu)}\tilde f_{stat}(M_b)~,
\eeq
with:
\beq
R_b(\mu)=\ga \alpha_s(\mu)\dr^{-\gamma_1/\beta_1}\Bigg{[} 1-\ga {\gamma_2\over \beta_1}-\gamma_1{\beta_2\over \beta_1^2}\dr a_s(\mu)\Bigg{]},
\eeq
where the two first coefficients of the $\beta$ functions are:
\beq
\beta_1=-{1\over 2} \ga 11-{2\over 3}n_l~\dr ,~~~~~~~~~\beta_2=-{1\over 4} \ga 51-{19\over 3}n_l\dr~.
\eeq
The {\it universal} coupling is connected to the physical decay constant as:
\beq
f_B\sqrt{M_B}=C_b(M_b)\tilde f_{stat}(M_b)+{\cal O}(1/M_b)~.
\label{eq:decayinfinite}
\eeq
It is also convenient to introduce the {\it universal} Renormalization Group Invariant (RGI) current and the associated coupling:
\beq
\hat J^\mu =R_b(\mu)\tilde J^\mu(\mu)~,~~~~~  \hat f_B=R_b(\mu) \tilde f_{stat}(\mu)~,
\label{eq:rgi_current}
\eeq
which we shall estimate in the following.
\section{HQET spectral sum rules for $\hat f_B$ in the static limit}
We shall be concerned with the  universal RGI 2-point-function\,\footnote{In HQET Lorentz structure is unimportant and it is only the parity which counts such that we shall omit it in the following.}:
\beq
\hat\Pi(q^2\equiv -Q^2)=i\int d^4x~e^{iqx}\la 0|\hat J(x)  \hat J^\dagger(0)  |0\ra
\eeq
 for determining the coupling $\hat f_B$ using QCD spectral sum rules (QSSR). 
Like in the case of the full theory, we can use either the Laplace (LSR) \cite{SVZ,SNRAF,BELL}:
\bea
{\cal L}(\tau,\mu)&\equiv&\lim_ {\begin{tabular}{c}
$Q^2,n\to\infty$ \\ $n/Q^2\equiv\tau$
\end{tabular}}
{(-Q^2)^n\over (n-1)!}{\partial^n \hat\Pi\over ( \partial Q^2)^n}\nnb\\
&=&\tau\int_{M_b^2}^{\infty}dt~e^{-t\tau}\frac{1}{\pi} \mbox{Im}\hat\Pi(t,\mu)~,
\label{eq:lsr}
\eea
or the $Q^2=0$ Moments sum rules (MSR) \cite{SVZ}:
\beq
{\cal M}^{(n)}(\mu)\equiv  {(-1)^n\over n!} {\partial^n \hat\Pi\over ( \partial Q^2)^n}\Big{\vert}_{Q^2=0}
=\int_{M_b^2}^{t_c}{dt\over t^{n+2}}~\frac{1}{\pi} \mbox{Im}\hat\Pi(t,\mu)~.
\label{eq:mom}
\eeq
 However, the use of the $Q^2=0$-moment sum rules is rather delicate as they do not have a proper infinite heavy quark mass limit.  We shall not consider these sum rules here \footnote{Some attempts to use $Q^2=0$-moment sum rules can be found in \cite{PENIN}.}. 
 For  the present analysis, it is convenient to introduce respectively the soft scale, the meson-quark mass-difference and the  HQET Laplace sum rule variable:
\beq
\omega={(q^2-M_b^2)\over  M_b}~,~~~\Delta M ={(M_B^2-M_b^2) \over M_b}~,~~~\tau_H=\tau M_b~,
\eeq
where $M_b$ is the on-shell quark mass and $\tau$ is the usual LSR variable used in the full QCD theory and has the dimension of GeV$^{-2}$. 
As usual, we parametrize the spectral function using the Minimal Duality Ansatz (MDA):
\beq
\frac{1}{\pi}\mbox{ Im}\hat\Pi(t)\simeq \hat f^2_B\delta(t-M^2_B)
  \ + \
  ``\mbox{QCD cont.}" \theta (t-t_c),
\label{eq:duality}
\eeq
where the accuracy for the sum rule approaches  has been explicitly  tested from heavy quarkonia systems in \cite{SNFB12}. The perturbative (PT) expression of the spectral function  related to the current $\tilde J^\mu(x)$ has been evaluated  to order 
{\cal O}$(\alpha_s) $ (NLO)  in \cite{DOSCH} and to {\cal O}$(\alpha_s^2) $ (N2LO) in \cite{CHET}. It reads:
\bea
{\rm Im}\tilde\Pi_{PT}(\omega)&=&{3\omega^2\over 8\pi}\Bigg{\{} 1+a_s(\mu) \ga {17\over 3} +{4\pi^2\over 9}+L_\omega\dr \nnb\\
&&+a_s^2(\mu)\Bigg{[} 99(15) +\ga {1657\over 72}+{97\pi^2\over 54}\dr L_\omega\nnb\\
&&+{15\over 8}L_\omega^2+n_l\Big{[}-3.6(4)\nnb\\
&&-\ga {13\over 12}+{2\pi^2\over 27}\dr L_\omega-{L_\omega^2\over 12}\Big{]}\Bigg{]}\Bigg{\}}~,
\label{eq:correl_HQET}
\eea
with: $L_\omega\equiv \ln(\mu^2/\omega^2) $ and $\hat \Pi_{PT}(\omega)\equiv R^2_b(\mu)\tilde \Pi_{PT}(\omega)$. We estimate the {\cal O}$(\alpha_s^3) $ (N3LO)
by assuming the  geometric growth of the PT series \cite{NZ} as a dual to the effect of a $1/q^2$ term \cite{CNZ,ZAK} which parametrizes the UV renormalon contributions.
NP corrections up to dimension 7 condensates has been obtained in \cite{GROZIN,BROAD2}. Including all the previous corrections, the sum rule reads for $M_b\to\infty$:
\bea
\hat f_B^2=e^{\tau_H\Delta M} R^2_b(\mu)
\Bigg{[}{ 1\over\pi}\int_0^{\omega_c}\hspace*{-0.2cm}d\omega~e^{-\omega\tau_H}
~{\rm Im}\tilde\Pi_{PT}\ga {\omega}\dr+NP\Bigg{]},
\label{eq:fhatb1}
\eea
where: $\omega_c={(t_c-M_b^2)/ M_b}$ and:
\bea
NP(\mu) &=&-\la\bar uu\ra(\mu) \Big{[}1+2a_s(\mu)-{M_0^2\over 4}\tau_H^2+{\pi\la\alpha_s G^2\ra\over 18}\tau_H^4\Big{]}\nnb\\
&&+\ga \pi \rho \la\bar uu\ra^2-{3g\la G^3\ra\over 356\pi^2}\dr{2\over 81}\tau_H^3~.
\label{eq:fhatb2}
\eea

\section{The QCD input parameters}
The QCD parameters which shall appear in the following analysis will be the  on-shell bottom quark mass $M_{b}$ (we shall neglect  the light quark masses $q\equiv u$ here and in the following),
the light quark condensate $\qq$,  the gluon condensates $ \lag
g^2G^2\rag\equiv \la g^2G^a_{\mu\nu}G_a^{\mu\nu}\ra$ and $\la g^3G^3\ra\equiv \la g^3f_{abc}G^a_{\mu\nu}G^b_{\nu\rho}G^c_{\rho\mu}\ra$, the mixed condensate $\mix\equiv {\la\bar qg\sigma^{\mu\nu} (\lambda_a/2) G^a_{\mu\nu}q\ra}=M_0^2\la \bar qq\ra$ and the four-quark 
 condensate $\rho\la\bar qq\ra^2$, where
 $\rho\simeq 2$ indicates the deviation from the four-quark vacuum 
saturation. Their values are given in Table {\ref{tab:param}}. We shall work with the running
light quark parameters known to order $\alpha_s^3$ \cite{SNB1,SNB2,RUNDEC}:
\bea
{\bar m}_{q,Q}(\mu)&=&
{\hat m}_{q,Q}  \ga-\beta_1a_s\dr^{-2/{
\beta_1}}\times C(a_s)
\nnb\\
{\la\bar qq\ra}(\mu)&=&-{\hat \mu_q^3  \ga-\beta_1a_s\dr^{2/{
\beta_1}}}/C(a_s)
\nnb\\
{\la\bar qg\sigma Gq\ra}(\mu)&=&-{M_0^2{\hat \mu_q^3} \ga-\beta_1a_s\dr^{1/{3\beta_1}}}/C(a_s)~,
\eea
${\hat m}_{q,Q}$ is the RGI quark mass, $\hat\mu_q$ is spontaneous RGI light quark condensate \cite{FNR}. The QCD correction factor $C(a_s)$ in the previous expressions is numerically:
\bea
C(a_s)&=& 1+0.8951a_s+1.3715a_s^2 +...~~{\rm :}~~ n_f=3~,\nnb\\
&=&1+1.1755a_s+1.5008a_s^2 +...~~{\rm :}~~ n_f=5~,
\eea
which shows a good convergence. 
We shall use:
\beq   
\alpha_s(M_\tau)=0.325(8) \lrar  \alpha_s(M_Z)=0.1192(10)
\label{eq:alphas}
\eeq
from $\tau$-decays \cite{SNTAU,BNP}, which agree perfectly with the world average 2012 \cite{BETHKE,PDG}: 
\beq
\alpha_s(M_Z)=0.1184(7)~. 
\eeq
 The value of the running $\la \bar qq\ra$ condensate is deduced from the value of $(\overline{m}_u+\overline{m}_d)(2)=(7.9\pm 0.6)$ MeV obtained in  \cite{SNmass,SNmass2} and the well-known GMOR relation: $(m_u+m_d)\la \bar uu+\bar dd\ra=-m_\pi^2f_\pi^2$. We shall use the value of the RGI  spontaneous mass to order $\alpha_s$ for consistency with the known $\alpha_s$ correction in the OPE:
\beq
\hat\mu_u=251(6)~{\rm MeV}.
\label{eq:lightmass}
\eeq
 The values of the running $\overline{MS}$ mass $\overline{m}_b(m_b)$  recently obtained in Ref. \cite{SNH10} from  bottomium sum rules, will also be used\,\footnote{These values agree and improve previous sum rules results \cite{SVZ,RRY,SNB1,SNB2,SNHmass,IOFFE}.}. Using the relation between the running $\overline{m}_b(\overline{m}_b)=4177(11)$ MeV from the $\Upsilon$-systems \cite{SNH10} and the on-shell (pole) $M_b$ masses (see e.g. \cite{SNB1,SNB2,CHET}, one can deduce to order $\alpha_s^2$:
 \beq
 M_b=4804(50)_{\alpha_s}~\to~\alpha_s(M_b)=0.2326(22)~,
 \label{eq:polemass}
\eeq
where the error is mainly due to the one of $\alpha_s$.
 This large error has to be contrasted with the precise value of the running mass,
and can be an obstacle for a precise determination of $\hat f_B$ from HQET at a given $\mu$.
On can see  in Section 8 that a direct extraction of the on-shell mass from the HQET at the same $\alpha_s^2$ order  leads to about the same value and error which is an (a posteriori) self-consistency check of the value and error used in Eq. (\ref{eq:polemass}) for the analysis. We are aware that
the inclusion of the known $\alpha_s^3$-correction and an estimate of the PT higher order terms using a geometric growth of the PT coefficients  \`a la Ref. \cite{NZ,CNZ} increase the value of $M_b$ by about $(100 \sim 200)$ MeV, which could only be considered if one works to higher order in $\alpha^n_s~(n\geq 3)$.  These large order terms are expected to be dual to the $1/q^2-$term which mimics the UV renormalon contribution. On the other, some eventual IR renormalon contributions are usually expected to be absorbed by the ones of the QCD condensates when the mass terms are included into the OPE. However,  the use of the pole mass  $M_b$ is only a convenient numerical step in our numerical analysis, as we could have worked from the beginning with the running mass. Therefore, our results on the running mass $\overline{m}_b$ truncated
at a given PT order are not (a priori) affected by the IR renormalon contributions.
{\scriptsize
\begin{table}[hbt]
\setlength{\tabcolsep}{0.2pc}
 \caption{
QCD input parameters. 
 }
    {\small
\begin{tabular}{lll}
&\\
\hline
Parameters&Values& Ref.    \\
\hline
$\Lambda(n_f=4)$& $(324\pm 15)$ MeV &\cite{SNTAU,BNP,BETHKE}\\
$\alpha_s(M_\tau)$& $0.325(8)$&\cite{SNTAU,BNP,BETHKE}\\
$\overline{m}_b(\overline{m}_b)$&$4177(11)$ MeV&average \cite{SNH10}\\
$\hat \mu_q$&$251(6)$ MeV&\cite{SNB1,SNmass}\\
$M_0^2$&$(0.8 \pm 0.2)$ GeV$^2$&\cite{JAMI2,HEID,SNhl}\\
$\la\alpha_s G^2\ra$& $(7\pm 1)\times 10^{-2}$ GeV$^4$&
\cite{SNTAU,LNT,NEUF,SNI,fesr,YNDU,SNHeavy,BELL,SNH10,SNG}\\
$\la g^3  G^3\ra$& $(8.2\pm 1.0)$ GeV$^2\times\la\alpha_s G^2\ra$&
\cite{SNH10}\\
$\rho \la \bar qq\ra^2$&$(4.5\pm 0.3)\times 10^{-4}$ GeV$^6$&\cite{SNTAU,LNT,JAMI2}\\
\hline
\end{tabular}
}
\label{tab:param}
\end{table}
}

\section{The LSR determination of the RGI $\hat f_B$ in the static limit}
\subsection*{\b Analysis of the convergence of the PT series}
\begin{figure}[hbt] 
\begin{center}
{\includegraphics[width=9cm  ]{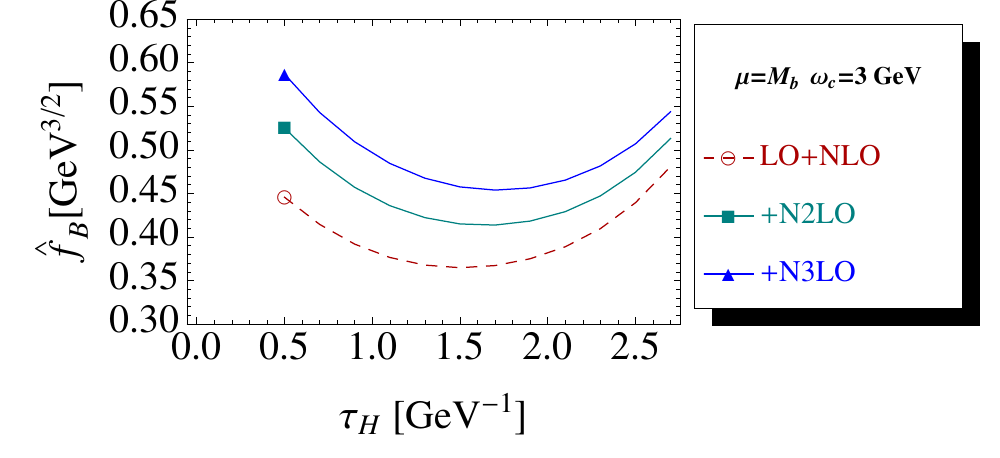}}
\caption{
\scriptsize 
 $\tau_H$-behaviour of $\hat f_B$  for $\mu=M_b$ and $\omega_c$= 3 GeV for different
truncations of the PT series . 
}
\label{fig:fhatconv}
\end{center}
\end{figure}

\nin
We study the effect of the truncation of the PT series on the value of  $\hat f_B$ from
Eqs. (\ref{eq:fhatb1}) and (\ref{eq:fhatb2}). For a given value of $\mu=M_b$ and $\omega_c$=3 GeV, we show the result of the analysis in Fig. \ref{fig:fhatconv}. At the minimum (optimal) value,  $\hat f_B$ moves from 0.365 (LO+NLO) to  0.414 (+N2LO) to 0.454 (+N3LO) GeV$^{3/2}$, i.e. a change of about 13\% from LO+NLO to N2LO and of about 8.8\% from N2LO to N3LO, which indicates a slow convergence of the PT series. We shall consider the N3LO contribution as a systematic
error from the truncation of the PT series.
\nin
\subsection*{\b Analysis of the $\tau_H$ and $\omega_c$ stabilities}
\begin{figure}[hbt] 
\begin{center}
\centerline {\hspace*{-5.3cm} a) }\vspace{-0.9cm}
{\includegraphics[width=9cm  ]{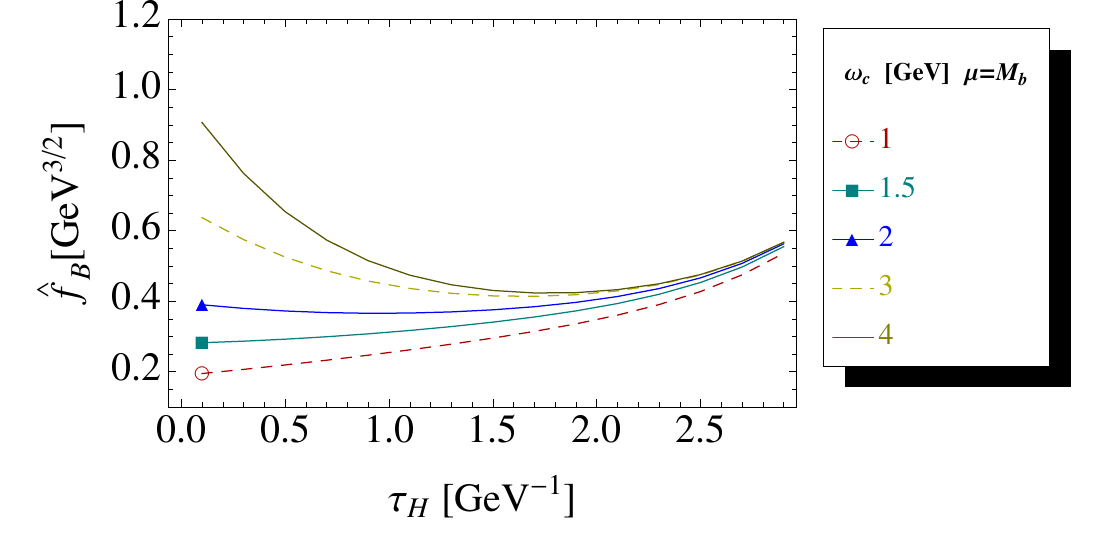}}
\centerline {\hspace*{-5.5cm} b) }\vspace{-0.6cm}
{\includegraphics[width=9cm  ]{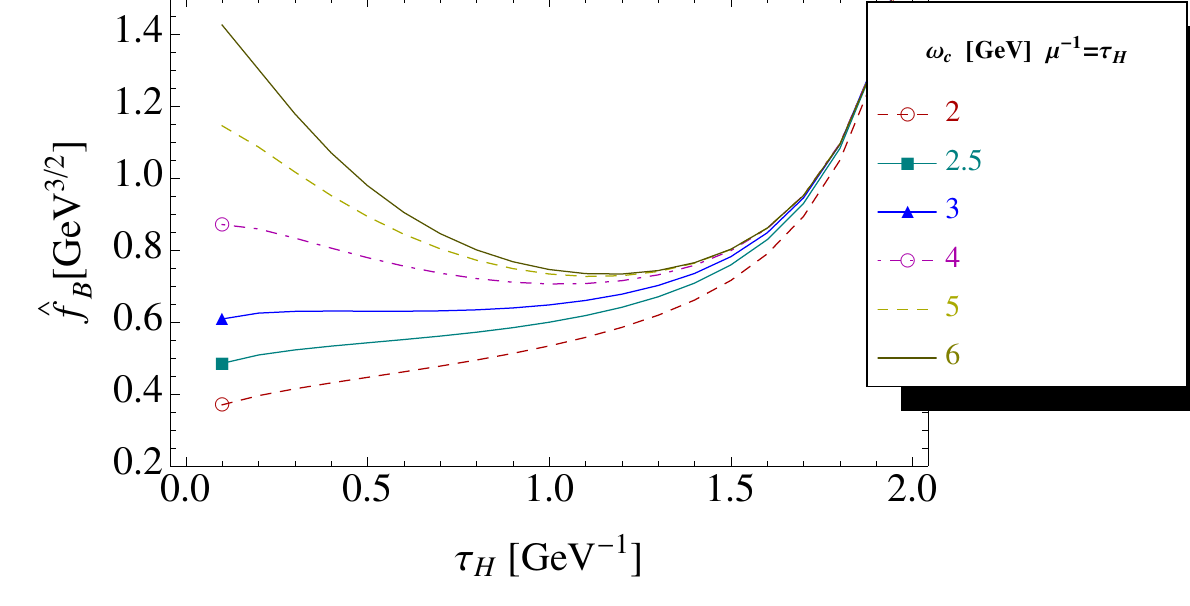}}
\caption{
\scriptsize 
{\bf a)} $\tau_H$-behaviour of $\hat f_B$  for $\mu=M_b$ and for different values of $\omega_c$; {\bf b)} The same as a) but for $\mu=\tau_H^{-1}$.
}
\label{fig:fhat_stab}
\end{center}
\end{figure}

\nin
We show in Fig. \ref{fig:fhat_stab} the $\tau_H$-behaviour of the result for a given value of $\mu$ 
and for different values of $\omega_c$ where the PT series is known to N2LO.  The $\tau_H$-stabilities are:
\beq
\tau_H\simeq (1~, 1.5) ~{\rm GeV}^{-1}~,
\label{eq:stability}
\eeq
where $\tau_H\simeq$ 1 GeV$^{-1}$ is obtained for $\omega_c \simeq 2$ GeV (beginning of $\tau_H$-stability), while $\tau_H\simeq$ 1.5 GeV$^{-1}$ corresponds to the beginning of the $\omega_c$-stability which is $\omega_c\simeq 4$ GeV. We consider, as optimal and conservative values, the ones obtained in the previous range of $\omega_c$ values. For, e.g $\mu=M_b$, we obtain in this way:
\beq
\hat f_B (M_b)= (0.373-0.427)~{\rm GeV^{3/2}}~.
\eeq
This range of values is much larger range than the one used in the current literature which appears to be an ad hoc choice.  
\nin

\subsection*{\b Summary of the results for $\hat f_B$ and error calculations}
\nin
At a given value of $\mu$, we estimate the errors induced by the QCD parameters compiled  in Table \ref{tab:param}. 
We summarize the results of the analysis in Table \ref{tab:fhat_res}.
{\scriptsize
\begin{table}[hbt]
\setlength{\tabcolsep}{0.4pc}
 \caption{
Central values and corresponding errors for $\hat f_B$ in units of $10^{-3}\times$GeV$^{3/2}$ from the LSR at different values of the subtraction point $\mu$ in units of GeV for $M_b=4804(50)$ MeV. The +(resp. --) sign means that the values of $\hat f_B$  
increase (resp. decrease) when the input increases (resp. decreases).  The total error comes from a quadratic sum.}
    {\small
\begin{tabular}{cccccccccc}
&\\
\hline
$\mu$&$\hat f_{B}$&$t_c$&$\alpha_s$&$\alpha_s^3$&$M_b$&$\la\bar uu\ra$&$\la G^2\ra$&$M^2_0$& \it Total\\
\hline
$\tau_H^{-1}$&489&+26&+10&+76&$-34$&+3&0&$-2$&88\\
1&509&+19&$+13$&+113&$-54$&+5&+2&$-22$&129\\
2&441&+24&$+6$&+63&$-40$&+5&+2&$-11$&80\\
3&418&+26&$+5$&+51&$-35$&+4&+1&$-8$&68\\
4&406&+26&$+4$&+44&$-33$&+5&+1&$-6$&62\\
$M_b$&400&+27&$+2$&+42&$-32$&+4&+1&$-7$&60\\
5&398&+28&$+4$&+40&$-33$&+5&+1&$-4$&60\\
6&392&+32&$+3$&+38&$-32$&+5&+1&$-4$&60\\
\hline
\end{tabular}
}
\label{tab:fhat_res}
\end{table}
}
\nin
\begin{figure}[hbt] 
\begin{center}
{\includegraphics[width=8cm  ]{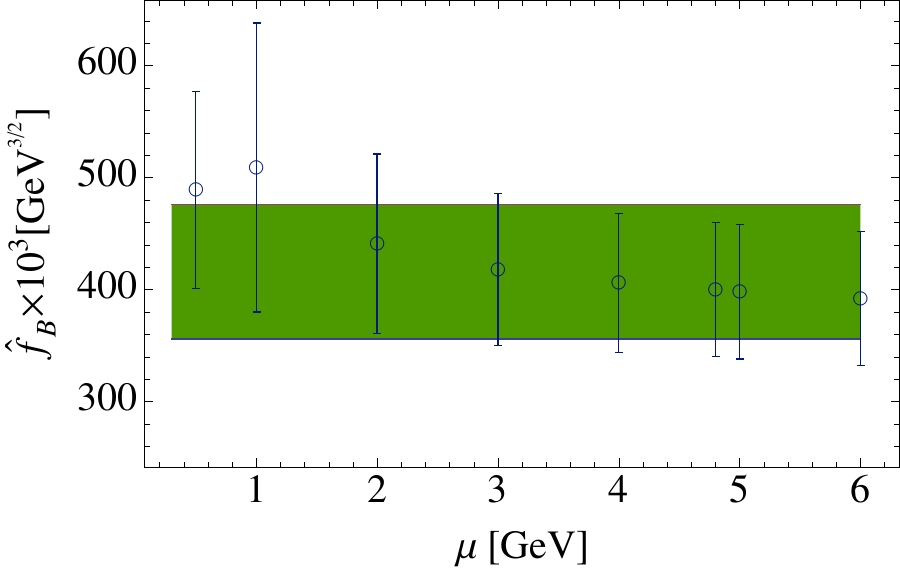}}
\caption{
\scriptsize 
 $\hat f_B$  at N2LO for different values of the subtraction point $\mu$ from Table \ref{tab:fhat_res}. The green coloured 
region corresponds to the one spanned by the average value and its corresponding error. 
}
\label{fig:fhat_data}
\end{center}
\end{figure} 
\nin
\nin
We show in Fig \ref{fig:fhat_data} the ``sum rules data points" at different values of the subtraction
point $\mu$. The error is large at small $\mu$ due to the bad behaviour of the PT series at low scale
which confirms the scepticism of the authors of Ref. \cite{BROAD} on the reliable extraction of $\hat f_B$ at a such low scale. However, as we have shown in previous section, the convergence of the PT series improves obviously at larger scale which enables to extract $\hat f_B$ with a reasonable
accuracy of about 16-15\% for $\mu\geq 3$ GeV. Fitting the previous data by an horizontal line or taking their average, we deduce the final value of the RGI universal coupling:
\bea
\hat f_B&=& 0.416(25)_{mean}(48)_{var}~{\rm GeV}^{3/2}\nnb\\
&=& 0.416(54)~{\rm GeV}^{3/2}~.
\label{eq:fbinv}
\eea
The estimate of the error is more delicate as there are not (a priori) any rigourous ways for obtaining it due to the  unclear eventual correlations among these different points. Here, we have deduced the error by adding the one 25 ${\rm GeV}^{3/2}$ from the weighted average (assuming uncorrelated data points) which is dominated by the most accurate prediction to the one 48 ${\rm GeV}^{3/2}$ obtained from the square root of the usual unbiaised estimator for $n-$number of  data points\,\cite{MUANZA}:
\beq
\Delta x=\sqrt{var(x)\equiv{1\over n-1}\sum_{i=1}^n(x_i-\la x\ra)^2}~,
\eeq
where $\la x\ra$ is the mean value. The size of the error is comparable with the one of about 60 ${\rm GeV}^{3/2}$ from the best determination at $\mu\approx M_b$.
A frequently used estimate of the error  induced by the truncation of the PT series would be obtained by varying the scale $\mu^{-1}$ from 1/2 to 2 times the value of $ \tau_H^0$ at which the sum rule is optimized instead of estimating the size of the $\alpha_s^3-$term like done above. Besides
the fact that this choice of range is arbitrary, the value of $ \tau_H^0$ at which the sum rule is optimized is quite large [see Eq. (\ref{eq:stability})] such that at $\mu^{-1}=2\tau_H^0$ the PT series breaks down rendering the approach inadequate. Instead, we can consider the value $\mu= (3.5\pm 2)$ GeV where the central value would correspond to the average obtained in Eq. (\ref{eq:fbinv}). 
Keeping only terms to order $\alpha_s^2$ and adding the different errors quadratically, one would obtain a final error of about 51 ${\rm GeV}^{3/2}$ which is slightly lower than the one in Eq.~(\ref{eq:fbinv}). Due to the arbitrariness of the choice of the range variation of $\mu$, we shall  only consider this result as an informative value and we shall not retain it in the final prediction. \\
Here and in the following, we shall alternatively estimate the final error which does not suffer from the previous drawbacks by taking the  one  coming from the most accurate measurement here at $\mu=M_b$. Then, we obtain:
\beq
\hat f_B= 0.416(60)~{\rm GeV}^{3/2}~,
\eeq
from which, we can deduce the value of the static coupling evaluated at, e.g., $M_B$=5.28 GeV:
\beq
\tilde f_{stat}(M_B)=0.603(2)_{\alpha_s}(87)_{\hat f_B}~{\rm GeV}^{3/2}~.
\label{eq:fbstat}
\eeq
The corresponding decay constant from Eq. (\ref{eq:decayinfinite}):
\beq
f_B^\infty=234 (1)_{\alpha_s}(33)_{\hat f_{B}}~\rm{MeV}~,
\label{eq:fbinf}
\eeq
which is relatively large compared to the value of $f_B=206(7)$ MeV obtained from the full QCD theory \cite{SNFB12} suggests some large $1/M_b$-corrections which we shall analyze in the next section.
We consider the previous results in Eqs. (\ref{eq:fbinv}) to (\ref{eq:fbinf}) as improvements of previous results in the literature~\cite{SNIMET,GROZIN,DOSCH,NEUBERT,ELETSKY,PENIN}.
Here, we have used updated values of the QCD input parameters. We have varied the continuum threshold and subtraction point $\mu$ in a larger range than in the existing literature in order to extract $\hat f_B$ for different values of the subtraction constant $\mu$. We have included NP contributions of higher dimensions ($d\leq 7$), though small, which are important for controlling the convergence of the OPE at a relatively large value of $\tau_H$ where the sum rule is optimized. We have also included in the error an estimate of the N3LO contribution based on a geometric growth of the PT series which controls the convergence of the PT series.
\section{$1/M_b$ corrections and value of the decay constant $ f_B$}
\nin
Taking into account the mass-difference between the meson $M_B$ and the on-shell quark 
mass $M_b$, the relation in Eq. ( \ref{eq:decayinfinite}) expressed in terms of the RGI coupling
in Eq. (\ref{eq:rgi_current}) becomes:
\beq
f_B^2=\ga {M_b\over M_B}\dr ^3\Bigg{[} {C_b(M_b)\over R_b(M_b)}{\hat f^2_B\over M_B} +\delta  f^2_B\Bigg{]}~,
\label{eq:fbphys}
\eeq
where : $M_B=5.279$ GeV and we shall use the value of $M_b$ in Eq. (\ref{eq:polemass}). 
\subsection*{\b LSR expression of $1/M_b$ correction $\delta f_B^2$ }
\nin
The $1/M_b$ corrections $\delta f_B^2$ to the HQET two-point correlator can be obtained by subtracting 
its expression in the full theory with the one of HQET in the limit $M_b\to \infty$. The $1/M_b$ 
correction to the physical decay constant $f_B$ reads (see e.g. \cite{PENIN}):
\bea
\delta  f^2_B= {e^{\tau_H\Delta M } \over M_B} \Bigg{[}{ 1\over\pi}\int_0^{\omega_c}\hspace*{-0.2cm}d\omega~e^{-\omega\tau_H}
~{\rm Im}\delta\Pi_{PT}\ga {\omega}\dr+\delta_{NP}\Bigg{]},
\label{eq:deltafb}
\eea
where:
\beq
{\rm Im}\delta\Pi_{PT}={\rm Im}\Pi_{PT}-C_b(M_b){R_b(\mu)\over R_b(M_b)}{\rm Im }\hat\Pi_{PT}~. 
\label{eq:deltafb1}
\eeq
Up to order $\alpha_s$, it reads:
\bea
{\rm Im}\delta\Pi_{PT}(x)&=&-{3\over 8\pi}{M_b^2}{x^3\over 1+x}\Bigg{\{}1+a_s\Bigg{[} \nnb\\
&&{1\over 2}\ga {13\over 4}+{\pi^2\over 3}-{3\over 2}\ln{x}\dr-{F(x)\over x}\Bigg{]}\Bigg{\}}\nnb\\ 
\eea
where : $x\equiv  {\omega/ M_b}$ and:
\bea
F(x)&=&2{\rm Li}_2(-x)+\ln(x)\ln(1+x)-{x\over 1+x}\ln(x)\nnb\\
&&+{1+x\over x}\ln(1+x)-1~.
\eea
The order N2LO $\alpha_s^2$ PT correction to the spectral function can be numerically obtained by subtracting the complete  expression in full QCD obtained in \cite{CHET} with the HQET asymptotic result in Eq. (\ref{eq:correl_HQET}) and by using the relation in Eq. (\ref{eq:deltafb}). In the same way, we estimate the N3LO PT corrections assuming a geometric growth of the PT series both in full QCD and HQET theories. 
The NP corrections read up to d=5 condensates \cite{DOSCH,PENIN}:
\bea
\delta_{NP}(\mu)&=&2a_s(\mu)\la \bar uu\ra (\mu)  \int_0^\infty {d\omega \over M_b}~{e^{-\omega\tau_H}\over 1+\omega / M_b}\nnb\\
&&+{\la \alpha_s G^2\ra\over 12\pi M_b}-\ga{\tau_H\over 2M_b}\dr{\la\bar qg\sigma Gq\ra}(\mu)~. 
\eea
\subsection*{\b Analysis of the convergence of the PT series }
\begin{figure}[hbt] 
\begin{center}
{\includegraphics[width=9cm  ]{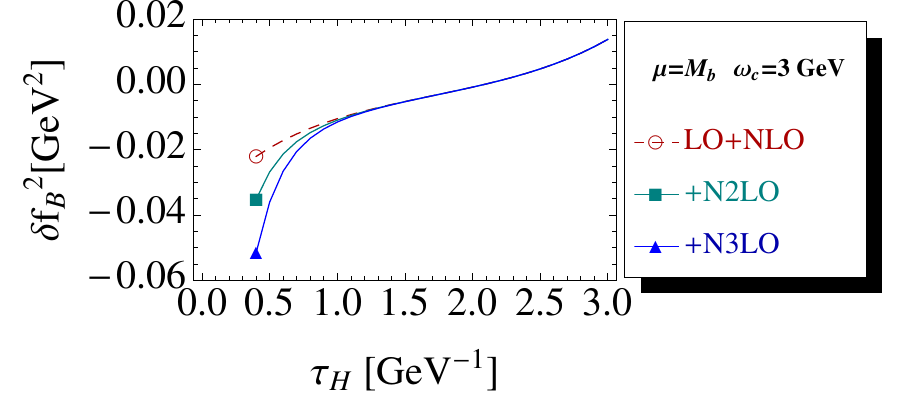}}
\vspace*{-0.5cm}
\caption{
\scriptsize 
$\tau_H$-behaviour of $\delta f^2_B$  for $\mu=M_b$ and $\omega_c$= 3 GeV for different truncations of the PT series where the contributions of condensates up to d=5 have been included. 
}
\label{fig:delta_conv}
\end{center}
\end{figure}

\nin
Like in the case of $\hat f_B$, we study the convergence of the PT series. We notice that the $\alpha_s^2$ and $\alpha_s^3$ corrections are very small for $\tau_H\leq$ 1 GeV$^{-1}$, which  can then be neglected.
The analysis
is shown in Fig. \ref{fig:delta_conv}. 
\nin
\subsection*{\b Analysis of the $\tau_H$ and $\omega_c$ stabilities}
\begin{figure}[hbt] 
\begin{center}
\centerline {\hspace*{-4.8cm} a) }\vspace{-0.6cm}
{\includegraphics[width=8cm  ]{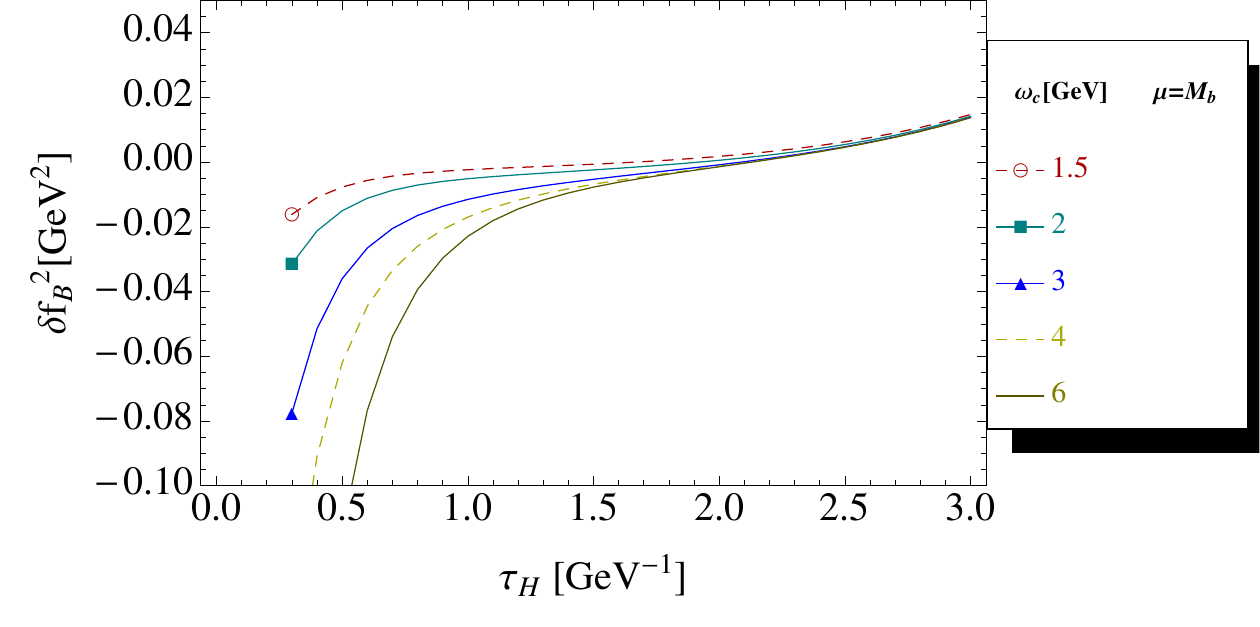}}
\centerline {\hspace*{-4.cm} b) }\vspace{-0.6cm}
{\includegraphics[width=8cm  ]{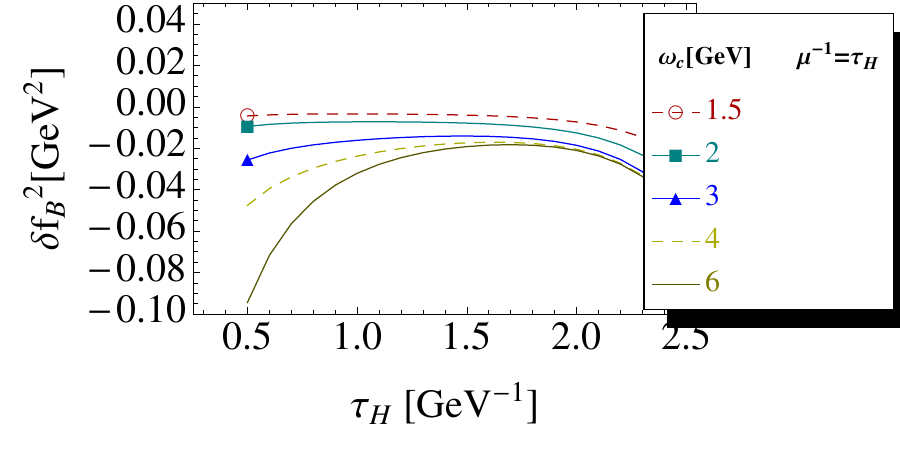}}
\vspace*{-0.2cm}
\caption{
\scriptsize 
{\bf a)} $\tau_H$-behaviour of $\delta f^2_B$  for $\mu=M_b$ and for different values of $\omega_c$ by  including the d=5 condensate in the OPE; {\bf b)} the same as a) but for $\mu=\tau_H^{-1}$.
}
\label{fig:delta_fb_tau}
\end{center}
\end{figure}

\begin{figure}[hbt] 
\begin{center}
{\includegraphics[width=9cm  ]{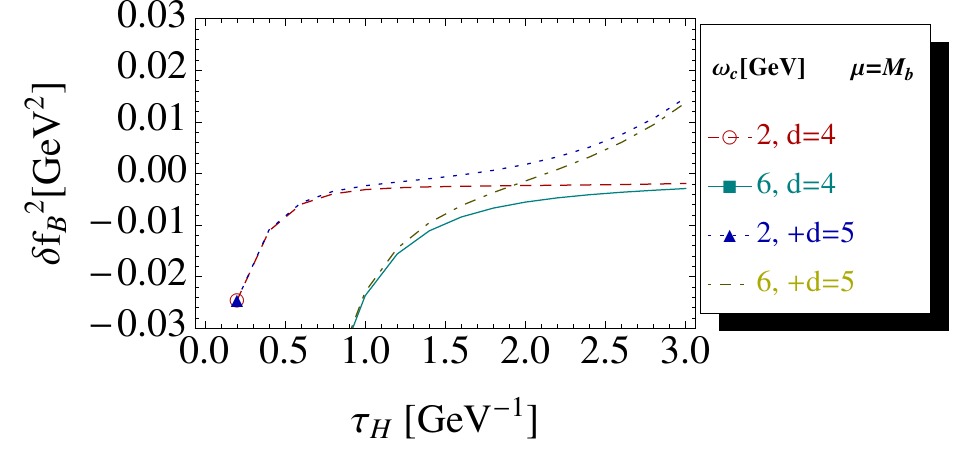}}
\vspace*{-0.5cm}
\caption{
\scriptsize 
$\tau_H$-behaviour of $\delta f^2_B$  for $\mu=M_b$ and $\omega_c$=  1.5 and 6 GeV for different trunctions of the OPE  by the inclusion of the d=4 condensate or by the inclusion of d=4+5 condensates. 
}
\label{fig:delta_cond}
\end{center}
\end{figure}

\nin
We show in Fig. \ref{fig:delta_fb_tau} the $\tau_H$-behaviour of $\delta f^2_B$ for different values of $\omega_c$ and including the $d=5$ condensates. 
\nin
We study the effects of the $d=5$ condensates on the $\tau_H$-stability for given  two extremal values  of $\omega_c$ (beginning of $\tau_H$ and of $\omega_c$-stabilites) . The analysis is shown in Fig. \ref{fig:delta_cond} from which we consider as optimal results the ones corresponding to the range:
\beq
\tau_H\simeq (1.6\sim 2.2)~{\rm GeV}^{-1}~,
\eeq
considering the fact that the inflexion point is not precisely localized. 
\nin
\subsection*{\b Results for $\delta f_B^2$}
\begin{figure}[hbt] 
\begin{center}
{\includegraphics[width=8cm  ]{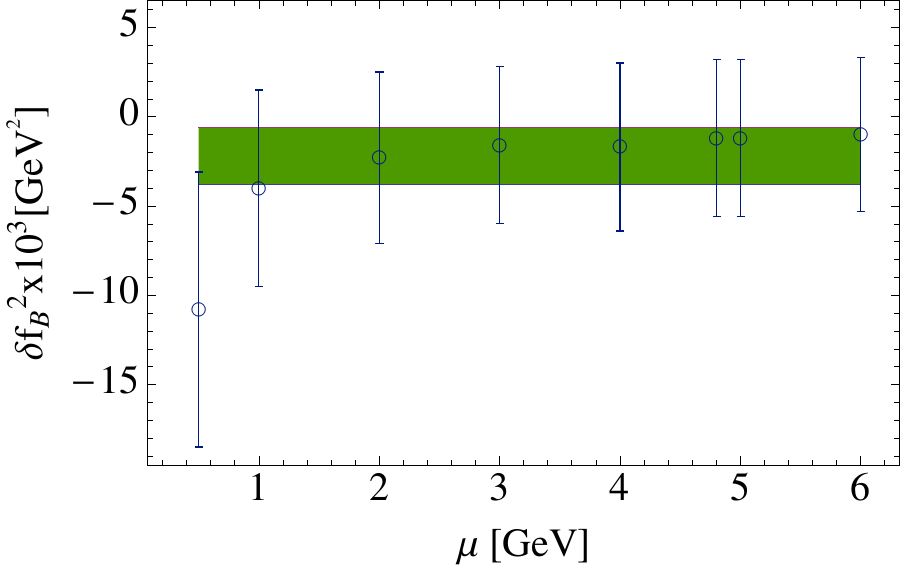}}
\caption{
\scriptsize 
``Optimal sum rules data" of $\delta f^2_B$  for different values $\mu$. The green coloured region is the one obtained from the average and its corresponding error.}
\label{fig:delta_fb_plot}
\end{center}
\end{figure}

\nin
The ``sum rules data" of $\delta f_B^2$ for different values of $\mu$ are shown Table \ref{tab:deltafb} and in Fig. \ref{fig:delta_fb_plot}, where the main errors come from the localization of $\tau_H$ from 1.8 to 2.2 GeV$^{-2}$ and the one induced from its corresponding $\omega_c$ values. The errors from the QCD parameters are negligible. 
{\scriptsize
\begin{table}[hbt]
\setlength{\tabcolsep}{1.2pc}
 \caption{
Central values and corresponding errors for $\delta f^2_{B}$ from the LSR at different values of the subtraction point $\mu$ and for $M_b=4804(50)$ MeV.}
    {\small
\begin{tabular}{ccc}
&\\
\hline
$\mu$ [GeV]&$-\delta f^2_{B}$ $\times10^{3}$[GeV$^2$]&Error $\times10^{3}$[GeV$^2$]\\
\hline
$\tau^{-1}_H$&$-10.8$&7.7\\
1&$-$4.0&5.5\\
2&$-$2.3&4.8\\
3&$-$1.6&4.4\\
4&$-$1.7&4.7\\
$M_b$&$-$1.2&4.4\\
5&$-$1.2&4.4\\
6&$-$1.0&4.3\\
\hline
\end{tabular}
}
\label{tab:deltafb}
\end{table}
}
\nin
Taking the average of different values, we deduce  the $1/M_b$ corrections due to $\delta f^2_B$:
\beq
\delta f_B^2=-2.2(1.6)\times 10^{-3}~{\rm GeV}^2~,
\label{eq:deltafB}
\eeq
where the error comes from the most accurate measurement at $\mu=\tau_H^{-1}$.
\section{$ f_B$ from HQET and from full QCD}
\nin
Combining the result in Eq. (\ref{eq:deltafB}) with the one in Eq. (\ref{eq:fbinv}) with the help of Eq. (\ref{eq:fbphys}), one obtains:
\bea
f_B^{hqet}&=&199(28.6)_{\hat f_B}(3)_{\delta f^2_B}(3)_{M_b}(0.3)_{\alpha_s}~{\rm MeV}\nnb\\
&=& 199(29)~{\rm MeV}~,
\label{eq:fbhqet}
\eea
where we have added the errors quadratically. Notice that, unlike the full QCD case \cite{SNFB12}, we have not tried to extract an upper bound on $f_B$ from the positivity of the spectral function because of the undefinite sign of $\delta f_B^2$ in Eq. (\ref{eq:fbphys}).   We can 
compare this result  with the one obained in the static limit in Eq. (\ref{eq:fbinf}), where one can see that the main corrections are to the $(M_b/M_B)^{3/2}$ ratio in Eq. (\ref{eq:fbphys}). This HQET 
result is consistent with the one from the average of LSR and Moment sum rules in full QCD~\cite{SNFB12}:
\beq
f_B^{qcd}=206(7)~{\rm MeV}~.
\label{eq:fbqcd}
\eeq
\section{Extraction of the $b$-quark mass  from HQET}
\begin{figure}[hbt] 
\begin{center}
\centerline {\hspace*{-5cm} a) }\vspace{-0.6cm}
{\includegraphics[width=8.8cm  ]{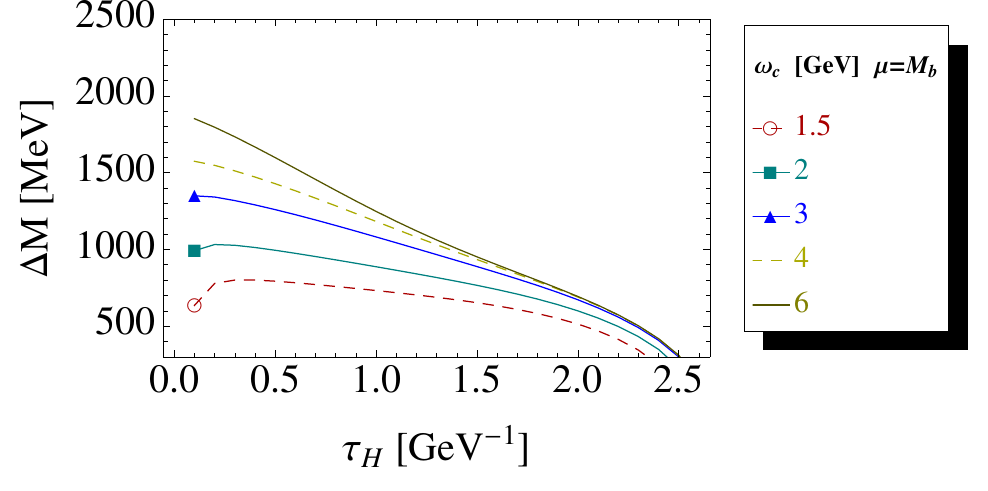}}
\centerline {\hspace*{-5cm} b) }\vspace{-0.6cm}
{\includegraphics[width=8.8cm  ]{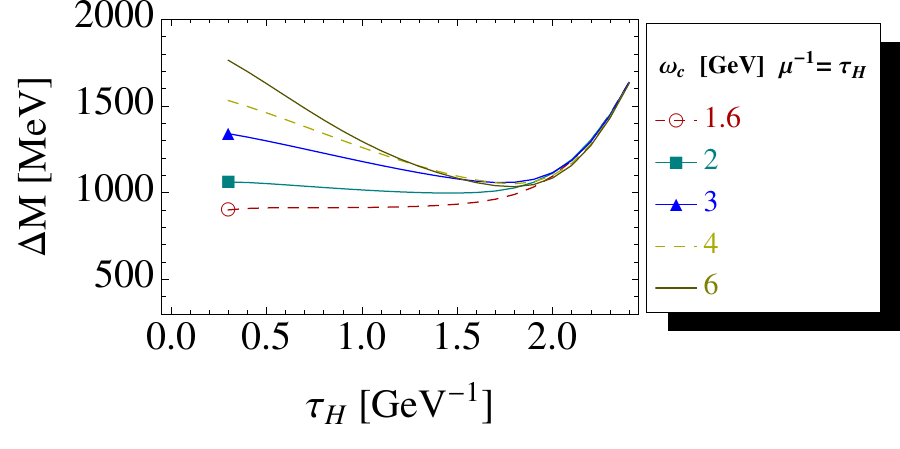}}
\vspace*{-0.5cm}
\caption{
\scriptsize 
{\bf a)} $\tau_H$-behaviour of $\Delta M$  for $\mu=M_b$ and for different values of $\omega_c$;
{\bf b)} the same as a) but for $\mu=\tau_H^{-1}$.
}
\label{fig:delta_tau}
\end{center}
\end{figure}

\begin{figure}[hbt] 
\begin{center}
{\includegraphics[width=8cm  ]{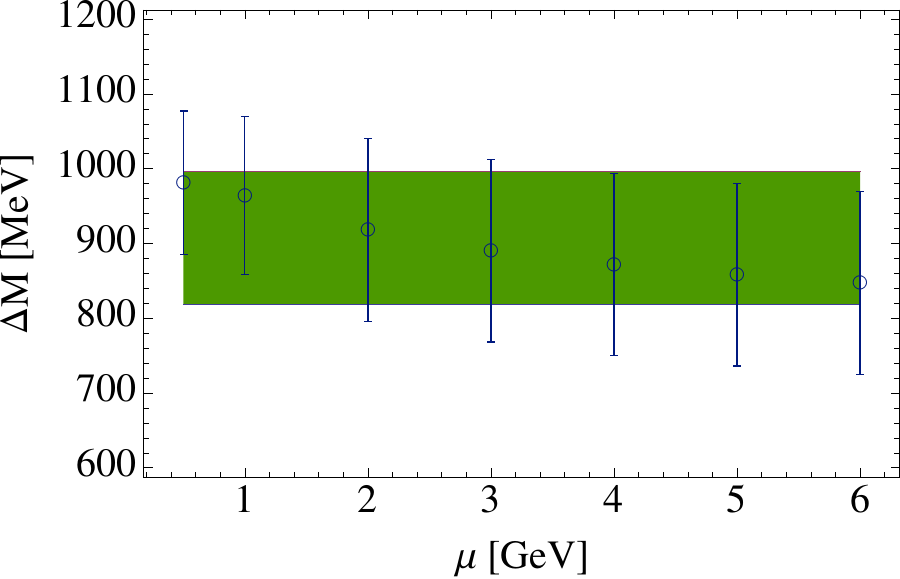}}
\caption{
\scriptsize 
``Optimal sum rules data" of $\Delta M$  for different values $\mu$. The green coloured region
corresponds to the mean value and its corresponding error.}
\label{fig:delta_plot}
\end{center}
\end{figure}

\nin
One can extract the meson-quark mass-difference $\Delta M$ using the ratio of the LSR obtained from
Eq. (\ref{eq:fbphys}):
\beq
{\cal R}_{\tau_H}\equiv {-{\partial\over \partial\tau_H} \ga\hat f_B^2 e^{-\tau_H \Delta M}\dr\over \hat f_B^2 e^{-\tau_H \Delta M}}=\Delta M\equiv {M_B^2-M_b^2\over M_b}~,
\eeq
where $M_b$ is the on-shell $b$-quark mass.  
We show the $\tau_H$-behaviour of $\Delta M$ in Fig. \ref{fig:delta_tau}. $\tau_H$-stability is obtained for $\tau_H$ about the values in Eq. (\ref{eq:stability}).
We show in Table \ref{tab:delta_res} the different sources of errors on $\Delta M$, where one can notice that the most important ones come from $\omega_c$, the estimated $\alpha_s^3$  and the mixed condensate contributions.
{\scriptsize
\begin{table}[hbt]
\setlength{\tabcolsep}{0.4pc}
 \caption{
Central values and corresponding errors for $\Delta M$ in units of MeV from the LSR at different values of the subtraction point $\mu$ in units of GeV. The +(resp. --) sign means that the values of $\Delta M$  
increase (resp. decrease) when the input increases (resp. decreases).  The total error comes from a quadratic sum.}
    {\small
\begin{tabular}{cccccccccc}
&\\
\hline
$\mu$&$\Delta M$&$t_c$&$\alpha_s$&$\alpha_s^3$&$M_b$&$\la\bar uu\ra$&$\la G^2\ra$&$M^2_0$& \it Total\\
\hline
$\tau^{-1}_H$&981&+69&+9&+64&+1&0&0&+18&96\\
1&964&+102&$+8$&+22&$+1$&$-6$&$-3$&$+15$&106\\
2&918&+109&$+8$&+44&$+1$&$-9$&$-4$&$+16$& 122\\
3&890&+109&$+7$&+51&$+1$&$-10$&$-4$&$+17$&122\\
4&872&+108&$+6$&+52&$+1$&$-11$&$-5$&$+17$&122\\
$M_b$&862&+108&+7&+54&+1&$-11$&$-5$&+16&123\\
5&858&+107&$+6$&+55&$+1$&$-12$&$-5$&$+16$&122\\
6&847&+106&$+6$&+57&$+2$&$-12$&$-5$&$+16$&122\\
\hline
\end{tabular}
}
\label{tab:delta_res}
\end{table}
}
\nin
We show in Fig. \ref{fig:delta_plot} the $\mu$-behaviour of different ``QSSR data points" from which we deduce the average:
\beq
\Delta M=907(89)~{\rm MeV}~,
\eeq
where the error comes from the most accurate measurement at $\mu=\tau_H^{-1}$.
 Using the previous value of $\Delta M$, one can extract the on-shell $b$-quark mass to order $\alpha_s^2$:
\beq
M_b^{hqet}= 4846(41)~{\rm MeV}~.
\eeq
Using the known relation between the on-shell and running quark mass to order $\alpha_s^2$ (see e.g. \cite{SNB1,SNB2,SNB3,RUNDEC})\,\footnote{We could have also extracted $\overline{m}_b(\overline{m}_b)$ directly by replacing the on-shell mass $M_b$  with $\overline{m}_b(\overline{m}_b)$ using their PT relation known to order $\alpha_s^2$, in the QCD expression. However, this procedure is not convenient in the numerical analysis.}, we deduce:
\bea
\overline{m}_b(\overline{m}_b)^{hqet}&=&4213(47)_{\alpha_s}(36)_{qssr}~{\rm MeV}\nnb\\
&=&4213(59)~{\rm MeV}~.
\label{eq:mbhqet}
\eea
\vspace*{-0.2cm}
\nin
\vspace*{-0.5cm}
\nin
\section{$\overline{m}_b(\overline{m}_b)$  from HQET and from full QCD}
\nin
The previous value of the running mass is in good agreement  with the one from heavy-light QCD spectral sum rules in full QCD to order $\alpha_s^2$ \cite{SNFB12}:
\beq
\overline{m}_b(\overline{m}_b)^{qcd}=4236(69)~{\rm MeV}~,
\label{eq:mbqcd}
\eeq
and with the more accurate result from the $\Upsilon$ sum rules to order $\alpha_s^3$ \cite{SNH10}:
\beq
\overline{m}_b(\overline{m}_b)^{\Upsilon}=4177(11)~{\rm MeV}~.
\label{eq:upsilon}
\eeq

{\scriptsize
\begin{table}[h]
\setlength{\tabcolsep}{0.6pc}
 \caption{
Results for  $f_B$ and $\overline{m}_b(\overline{m}_b)$ in units of MeV and comparison with  lattice simulations
using $n_f=2$ \cite{ETM,HEITGER} and $n_f=3$ \cite{HPQCD,MILC} dynamical quarks. $f_P$ are normalized as $f_\pi=130.4$ MeV. }
    {\small
\begin{tabular}{llll}
&\\
\hline
Observables& Methods &Refs.    \\
\hline
\boldmath${f_B}$\\
{\it QSSR}\\
$199(29)\equiv 1.53(23)f_\pi$&HQET &(this work)\\
$206(7)\equiv 1.58(5)f_\pi$&full QCD &\cite{SNFB12}\\
$\leq 235.3(3.8)\equiv 1.80(3)f_\pi$& full QCD& \cite{SNFB12}\\
{\it Lattice}\\
197(10)& ETMC&\cite{ETM}\\
 193(10)& ALPHA &\cite{HEITGER} \\
190(13)& HPQCD &\cite{HPQCD}\\
197(9)& FNAL& \cite{MILC}\\
\\
\boldmath$\overline{m}_b(\overline{m}_b)$&\\
{\it QSSR}\\
4213(59)& $B$-meson - HQET &(this work) \\
4236(69)& $B$-meson - full QCD  &\cite{SNFB12} \\
4177(11)&$\Upsilon$ - full QCD &\cite{SNH10}\\
{\it Lattice}\\
4290(140)&ETMC &\cite{ETM}\\
\hline
\end{tabular}
}
\label{tab:result}
\end{table}
}
\vspace*{-0.5cm}
\section{Summary and conclusions}
\nin
We have re-estimated $f_B$ and $M_b$ from HQET Laplace spectral sum rules to order $\alpha_s^2$ by including an estimate of the $\alpha_s^3$ and non-perturbative terms up to dimension $d=7$ condensates. We have also taken larger ranges of $\omega_c$, $\tau_H$ 
and $\mu$ values for extracting our optimal results. Most of these analyzes have not been done in previous literature \cite{GROZIN,DOSCH,NEUBERT,ELETSKY,PENIN}. Our results in Eqs. (\ref{eq:fbhqet}) and (\ref{eq:mbhqet}) are in good agreement with the ones from full QCD in Eqs. (\ref{eq:fbqcd}) and (\ref{eq:mbqcd}). 
These results are comparable with some other $\Upsilon$ sum rule determinations \cite{UPSILON} and with lattice results including $n_f=2$ or 3 dynamical quarks compiled in
Table \ref{tab:result} \cite{ROSNER,ASNER,ETM,HEITGER,HPQCD,MILC}.

\end{document}